\RequirePackage{fix-cm}
\documentclass[twocolumn]{svjour3}          

\smartqed  

\usepackage{graphicx}
\usepackage{multirow}
\usepackage{morefloats}
\usepackage{natbib}

\journalname{Celestial Mechanics and Dynamical Astronomy}

\begin{document}

\title{Characterizing Multi-planet Systems with Classical Secular Theory }

\author{Christa Van Laerhoven
	\and
	Richard Greenberg
}


\institute{C. Van Laerhoven \at
              Department of Planetary Science, University of Arizona, 1629 E. University Blvd., Tucson AZ, 85721 \\
              Tel.: 520-621-1594\\
              Fax: 520-621-4933\\
              \email{cvl@lpl.arizona.edu}
           \and
           R. Greenberg \at
              Department of Planetary Science, University of Arizona
}

\date{Submitted to Celestial Mechanics and Dynamical Astronomy: 08/22/11, revised 01/26/12, revised 02/08/12}

\maketitle

\begin{abstract}
Classical secular theory can be a powerful tool to describe the qualitative character of multi-planet systems and offer insight into their histories. The eigenmodes of the secular behavior, rather than current orbital elements, can help identify tidal effects, early planet-planet scattering, and dynamical coupling among the planets, for systems in which mean-motion resonances do not play a role. Although tidal damping can result in aligned major axes after all but one eigenmode have damped away, such alignment may simply be fortuitous. An example of this is 55 Cancri (orbital solution of Fischer et al., 2008) where multiple eigenmodes remain undamped.  Various solutions for 55 Cancri are compared, showing differing dynamical groupings, with implications for the coupling of eccentricities and for the partitioning of damping among the planets. Solutions for orbits that include expectations of past tidal evolution with observational data, must take into account which eigenmodes should be damped, rather than expecting particular eccentricities to be near zero.  Classical secular theory is only accurate for low eccentricity values, but comparison with other results suggests that it can yield useful qualitative descriptions of behavior even for moderately large eccentricity values, and may have advantages for revealing underlying physical processes and, as large numbers of new systems are discovered, for triage to identify where more comprehensive dynamical studies should have priority.

\end{abstract}

\section{Introduction}
\label{intro}

The acceleration of discoveries of extrasolar systems is providing a base of data for inferring the current properties as well as the histories of planetary systems. Understanding planetary systems in general provides the essential context for our own solar system as well. A system's current dynamical state is the result of the initial planet formation, orbital evolution in the nebula, close planet-planet encounters, and subsequent long-term tidal evolution. Thus its orbital characteristics and distributions provide a key to the systems' origin and evolution. An important step is to find ways to characterize systems that shed light on their history.

An essential part of the initial discovery process is determination of the current Keplerian orbital elements. However, mutual perturbations cause these elements to change on timescales short compared with the age of a system. Observers have long been well aware of this issue, so numerical integration is typically part of the orbit-fitting process because it helps rule out ranges of masses and orbital elements that might be consistent with observational data, but inconsistent with long-term stability. Mutual interactions are enhanced by orbital resonances, which are thus especially interesting from the point of view of dynamical theory. However, even in systems with no resonances, secular interactions can be important, and the nature of those interactions can be a key to the history and properties of these systems.

Keplerian orbital elements are constants of integration in the two-body problem, but they are not constant when multiple planets interact. In secular interactions, planets exchange angular momentum, so eccentricities and longitudes of pericenter vary periodically. As long as eccentricities are not too large and no mean-motion resonances are involved, the analytical solutions of classical secular theory provide constants of integration that represent the system better than the Keplerian elements. While in principle numerical integrations can represent the behavior more precisely, secular theory provides a way to characterize, classify, interpret, and compare the dynamical states of various systems. The ``character'' of a system as used here refers to any of a number of different qualitative behaviors revealed by secular theory. The extent to which a system exhibits certain characteristics defines what we mean by its current dynamical state.

In this paper we demonstrate how this characterization using classical, second-order secular theory can provide insight into the origin and evolution of the system. We use several planetary systems to illustrate various points, generally selecting a particular orbital solution (e.g. a published best-fit) to observations as an example model for this purpose. In this sense we are following in the spirit of \citet{WuGoldreich02}, who considered the unconfirmed (and ultimately not-existent) HD 83433 system and \citet{Mardling07}, who considered the HD 209458 system with an ad hoc extra outer planet. Both of those studies used idealized systems to demonstrate important aspects of secular behavior.  This paper too is not intended to be an exhaustive study of the systems discussed here, so the range of uncertainties in the planets' orbits are not considered. Similarly, many discovery papers on new planetary systems include discussions of the dynamical state of their best-fit orbits. In those discussions, as here, a full investigation of all possible cases within the range of observational uncertainty is beyond their scope. 

We consider several specific planetary systems and types of behavior to demonstrate how the classical treatment of secular theory can provide evidence for past events and evolution.  The key to the value of this approach is that classical secular theory can be described by sets of straightforward linear differential equations, so that the solutions are expressed as sums of eigenmodes.  The constants of integration that characterize the system are the magnitudes and phases of these eigenmodes, rather than Keplerian elements like the eccentricities and pericenter longitudes.  With understanding of these results, there is the potential to infer significant aspects of the past evolution by inspection of the values of these parameters.

One of the evolutionary processes that governs long-term evolution is the effect of tides. In a single-planet system, the planet's energy and angular momentum change, affecting the semi-major axis and eccentricity values (for example, \citealt{Jackson08}). The inclination is affected as well, but in this paper we consider only systems in which all orbital and rotational axes are (or can reasonably be assumed to be) aligned. In multi-planet systems, secular interactions redistribute angular momentum among the planets, so that the effect of tides is shared. Over the long term, the result is not only a change in Keplerian elements, but also a gradual qualitative change in the secular character of a system.

Notably, tidal evolution can lead to an alignment of the orientation of the major axes: In the presence of secular interactions, any process (such as tides) that tends to damp the eccentricity of one or more planets will actually have the effect of damping all of the eigenmodes, each at a different rate.  Eventually, when only the longest lived eigenmode remains, the major axes are aligned, a distinctly recognizable orbital behavior that is the hallmark of eccentricity damping. However, major axes being locked in alignment does not necessarily mean that only one eigenmode is dominant or that there has been much eccentricity damping \citep{GrVL61Vir}. Moreover, although one might anticipate that the eigenmode that dominates the behavior of the damped planet is the eigenmode that damps fastest, such is not necessarily the case.

The current secular character of a system might also be indicative of planet-planet scattering.  Suppose two planets undergo a close encounter, resulting in large eccentricities (including the possible ejection of one from the system as in the case described by \citet{FordRasio05}).  Immediately after the event, the other planets may retain the nearly circular orbits acquired during their formation within a nebula.  Then, the subsequent secular interactions must result in periodic changes in all the eccentricities, so that at any observational epoch, all the Keplerian eccentricities may be substantial.  However, because the secular interactions are periodic, the dynamical behavior in such a case is characterized by periodic returns to circular orbits, as discussed by \citet{BarnesGreen06a}. This condition is readily identified by inspection of the magnitudes of the eigenmodes for each planet. If a combination can sum to zero, it means that eccentricities must periodically return to zero.  Thus the secular solution can identify candidate systems that may have suffered a planet-planet scattering event, as discussed further in Sect.~\ref{55CncF08}.

The degree to which planets share the effects of the same eigenmodes can also be used to describe how strongly coupled their behavior is, and thus can shed light on (and help quantify) the closeness of packing of the planets, with important implications for the origin and evolution of the system.  The ability to quantify the strength of the coupling is also useful for checking the plausibility of observational orbit fits, as demonstrated for the case of 55 Cancri in Sect.~\ref{55Cnc}.

These and other examples of the insight that can be derived from classical secular theory are discussed below.  The underlying theory is comprehensively described in various textbooks (e.g. \citealt{BrouwerClemence}, Murray and Dermott 2000\cite{MurrayDermott}), and we review it briefly here (Sect.~\ref{sec2}) to make the presentation complete, to establish our notation, and to include the effect of eccentricity damping.


\section{Secular Behavior}
\label{sec2}


\subsection{Classical Theory}
\label{sec2pt1}

Classical secular theory involves averaging the disturbing force at a planet due to each other planet over the planets' orbital periods. Then the remaining disturbing potential is used to calculate the changes in Keplerian orbital elements over time. In effect, the gravitational interactions are computed as if each planet's mass is smeared out along its Keplerian orbit with the portion of the mass in any segment of this path determined by how much time the planet would spend there in unperturbed Keplerian motion. In truth, the disturbing potential is expanded in a Fourier series, and those terms are ignored that explicitly contain orbital longitudes in the arguments of the sines or cosines (\citealt{BrouwerClemence}, \citealt{MurrayDermott}). Accordingly, this approach would not be valid where any of the planets are in mean-motion resonance.

Here we also assume that the orbits are co-planar and that the eccentricities are small enough that terms in the disturbing function higher than second order in the eccentricities can be ignored. (See \citealt{Mardling10} for a discussion of tidal evolution in non-coplanar systems.) It is also quite straightforward to include General Relativity, because its effect on orbital precession has the same functional form as terms in the planetary disturbing function.

The variation equations take on a simple linear form if we replace the Keplerian elements eccentricity ($e$) and longitude of pericenter ($\varpi$) with the elements $h$ and $k$:

\begin{equation}
k_p = e_p \cos \varpi_p
\end{equation}

\begin{equation}
h_p = e_p \sin \varpi_p
\end{equation}

\noindent where the subscript $p$ is an integer denoting the planet, in order from nearest to farthest from the star. $h_p$ and $k_p$ are the Cartesian components of the eccentricity vector, whose magnitude is $e_p$ and direction is given by $\varpi_p$. Expressing Lagrange's equations for planetary perturbations in terms of $h$ and $k$ yields the linear differential equations:

\begin{equation}\label{kdot}
\dot{k_p} = - \sum_{j=1}^N A_{pj} h_j
\end{equation}

\begin{equation}\label{hdot}
\dot{h_p} = \sum_{j=1}^N A_{pj} k_j
\end{equation}

\noindent where $N$ is the number of planets in the system and the matrix $A$ depends on the masses and semi-major axes of the planets and the mass of the star as given, for example, by \citet{MurrayDermott}. Additional precession effects due to General Relativity (included in the calculations for this paper), oblateness of the primary, etc., can be accommodated by incorporation in $A_{11}$, $A_{22}$, etc. 

As standard first-order linear differential equations, the solution of Eqns.~\ref{kdot} and \ref{hdot} is a sum of eigenmodes:

\begin{equation}\label{ksol}
k_p = \sum_{m=1}^N E_m V_{mp} \cos (g_m t + \delta_m ) = \sum_{m=1}^N e_{mp} \cos (g_m t + \delta_m )
\end{equation}

\begin{equation}\label{hsol}
h_p = \sum_{m=1}^N E_m V_{mp} \sin (g_m t + \delta_m ) = \sum_{m=1}^N e_{mp} \sin (g_m t + \delta_m )
\end{equation}

\noindent where $V_{m}$ are the normalized eigenvectors (for a given $m$, the sum of the squares of the $N$ planets' components ($V_{mp}$) is unity) and $g_m$ are the corresponding eigenfrequencies of the matrix $A$. Here we let the subscript $m$ denote the eigenmodes in order of decreasing $g_m$. For a given eigenmode, the eigenvector components $V_{mp}$ describe how the amplitude of that mode is partitioned among the planets. The mode amplitudes $E_m$ and phases $\delta_m$ are the constants of integration, while the Keplerian elements vary with time. The constants of integration are determined by using the known eccentricities $e_p$, and longitudes of pericenter $\varpi_p$ (equivalent to the known $h_p$,$k_p$ values) at any given time (i.e. initial conditions).

\begin{figure*}
\includegraphics{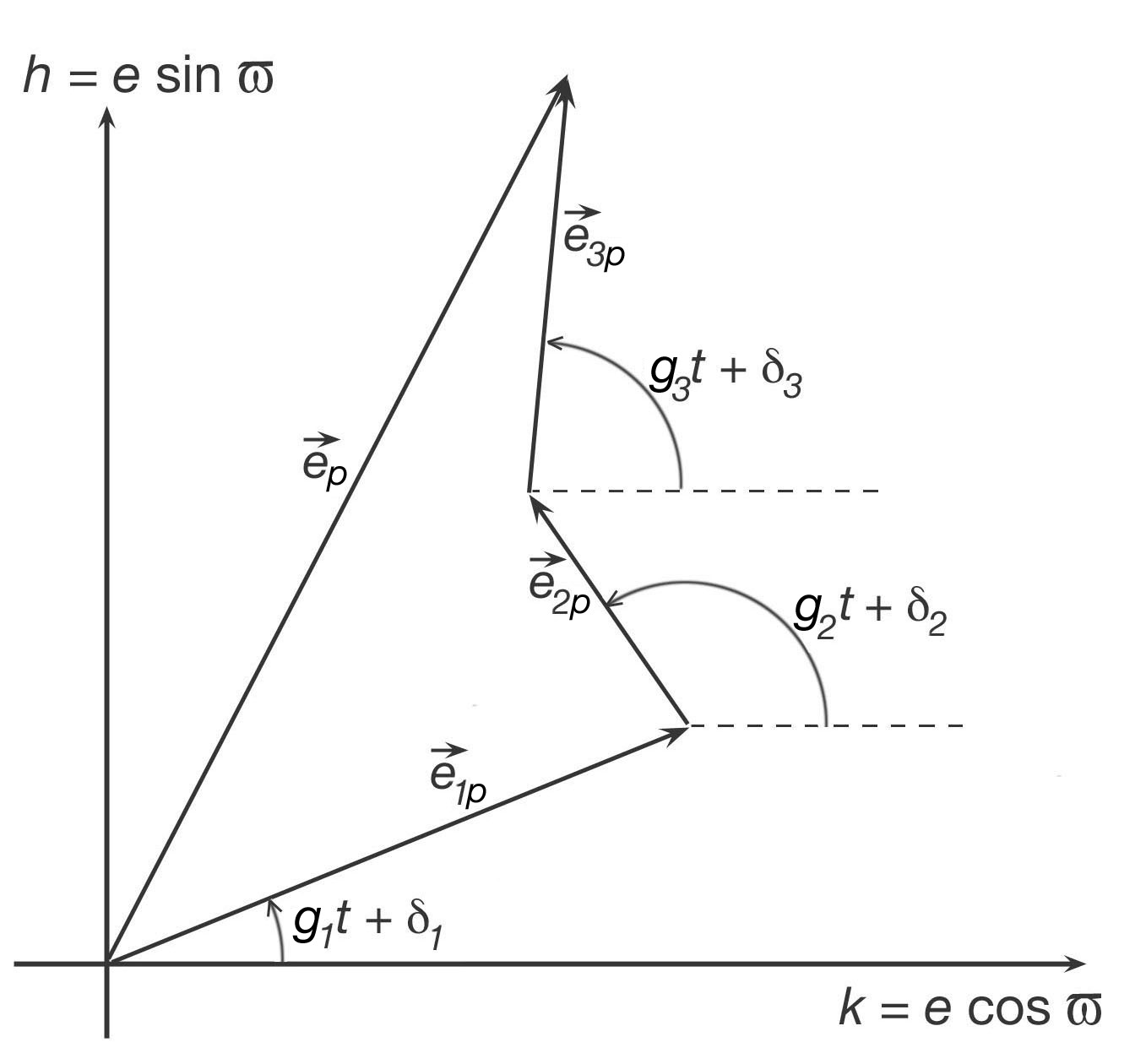}
\caption{
The behavior of the eccentricity of planet p according to the secular solution (Sect.~\ref{sec2}). The contribution of each mode to this planet is $e_{mp}=E_mV_{mp}$, and it rotates at a rate given by the corresponding eigenfrequency, $g_m$. The phase $\delta_m$ and the amplitudes of the modes $E_m$ are set by initial conditions.
}
\label{fig1}
\end{figure*}

Inspection of Eqns.~\ref{hsol} and \ref{ksol} shows that for each planet the ($h_p$,$k_p$) vector (which we call the eccentricity vector or $e$ vector, with length $e_p$ and direction $\varpi_p$) is the sum of $N$ component vectors in ($h$,$k$) space, each corresponding to a particular eigenmode and rotating in ($h$,$k$) space at an angular velocity given by $g_m$ (Figure~\ref{fig1}). Moreover, for a given eigenmode, the component vector contributed to the planets are all either parallel or anti-parallel, because they all point in the direction $g_m t+\delta_m$ or $g_m t+\delta_m+\pi$. So, if only one of the eigenmodes has non-negligible strength, all the planets' directions of pericenter would be aligned (if the signs of $V_{mp}$ are the same) or anti-aligned (if the signs of $V_{mp}$ are opposite). Equivalently, we can say the planets' major axes (or lines of apsides) are aligned or anti-aligned. Additional background on the basic geometry of secular behavior in orbital element space, e.g. how secular behavior may yield libration or circulation about alignment or anti-alignment of major axes, may be found in various references (e.g. Chapter 7 of \citealt{MurrayDermott}; \citealt{ChiangMurray}; \citealt{BarnesGreen06a,BarnesGreen06b}; \citealt{GrVL61Vir}).

For each mode, the relative magnitudes of the components $V_{mp}$ describe how strongly that mode affects a given planet compared with the other planets. And for a given planet, the relative magnitudes of the product $E_m V_{mp}$ describes which modes dominate that planet. These considerations can be used to determine how strongly coupled the interactions may be among various subsets of the planets in a system, as discussed in specific examples in the following sections.

Another useful point is that, like the matrix $A$, the eigenvectors and eigenfrequencies depend only on the masses and semi-major axes of planets in the system, and the mass of the star. Thus some understanding of the character of the system can be developed even if observations are insufficient to determine the eccentricity values precisely.


\subsection{Secular Behavior with Eccentricity Damping}
\label{sec2pt2}

Suppose some process (in addition to the gravitational effects taken into account in the derivation of Eqns. \ref{ksol} and \ref{hsol}) acts to damp the eccentricity of the inner planet according to

\begin{equation}\label{edampeqn}
\frac{de_1}{dt} = - F e_1
\end{equation}

\noindent where $F$ is positive and constant. Such eccentricity damping can be one effect of tidal dissipation (e.g. \citealt{GoldreichSoter1966}, \citealt{YoderPeale81}, \citealt{Jackson08}, \citealt{Matsum10}), although here we are not considering other effects of tides such as changes in the semi-major axis. Eccentricity damping can be incorporated into the secular theory by adding a term $-Fh_1$ to $dh_1/dt$ and $-Fk_1$ to $dk_1/dt$ in Eqns.~\ref{kdot} and \ref{hdot}, and therefore adding the term $-Fi$ to $A_{11}$ (where $i = (-1)^{1/2}$). The secular equations are still linear, so solutions are still straightforward (e.g. \citealt{ChiangMurray}). However, the eigenvalues will now be complex, which introduces an exponential decrease of each of the mode strengths, $E_m$, at a rate proportional to $F$. The damping coefficients for the various eigenmodes depend only on masses and semi major axes as they come from the diagonalization of the matrix $A$. So, which mode damps the fastest does not depend on the excitation of the mode. Consequently, $(dE_m/dt)/E_m$ is constant with time in this solution.

Even if the damping process acts directly on only one planet, the secular interactions redistribute angular momentum among the planets, so that eventually all the eccentricities die away. However, each eigenmode damps according to its own exponential timescale. Thus, as a damping proceeds a stage is reached where only the longest-lived eigenmode remains. This condition has been called a ``fixed point'' solution or, in recognition of the fact that eventually even the longest-lived eigenmode damps away, a ``quasi-fixed point solution'' \citep{WuGoldreich02,Mardling07,Batygin09}. As long as the one longest-lived eigenmode remains, this condition is characterized by aligned or anti-aligned major axes. Before this condition is reached, but after the other eigenmodes have become smaller than the long-lived mode, their contribution to the total solution of the behavior of the eccentricity vectors is so small that it can only produce librations around the alignment of major axes.

The various properties of the classical solution for secular behavior reviewed in this section have implications for the interpretation of the characteristics of observed extrasolar systems, as demonstrated by the following examples.


\section{Case Study Based on 55 Cancri}
\label{55Cnc}

The 55 Cancri system consists of five known planets: a ``super-Earth'' and four giant planets \citep{Fischer08,DawsonFabrycky10}). Like the terrestrial planets in our own solar system, the relatively small super-Earth is on an orbit interior to that of the giants. Unlike our solar system, the inner planet is close enough to its host star that tides may play a role in the planet's orbital evolution. Table~\ref{F1} shows the best fit to the radial velocity data for 55 Cancri from the discovery paper by \citet{Fischer08}. Here, to simplify notation, we refer to these planets by integer in order of increasing semi-major axis rather than by letters in order of discovery. More recent solutions by \citet{DawsonFabrycky10} take aliasing effects into account, yielding a much smaller orbit for the innermost planet. This shorter period has recently been confirmed by transit observations by \citet{Winn11} with the Microvariability and Oscillations of STars (MOST) telescope and \citet{Demory11} with the Spitzer Space Telescope.

Here, we consider three different solutions for the orbits in this system as examples to illustrate the value of classical secular theory for describing the dynamical character of a planetary system and identifying its implications. The three systems we consider below are the best-fit from \citet{Fischer08} and two of the solutions presented by \citet{DawsonFabrycky10}. Note that in these solutions planets 2 and 3 are near a 3:1 mean-motion resonance, which for our purposes of demonstration can be ignored. In all of these solutions several planets have strikingly similar longitudes of pericenter ($\varpi$), which can be the consequence of eccentricity damping as discussed in Sect.~\ref{sec2pt2}. Given that the inner planet is close to its star, one might expect tidal dissipation to have been responsible. However, before accepting such a conclustion we must consider the actual magnitude of each eigenmode and how each eigenmode affects the eccentricities of the planets.


\subsection{The First Orbit Solution}
\label{55CncF08}

In the orbit fit by \citet{Fischer08} (our Table~\ref{F1}), the super-Earth and three giant planets all lie within 1 AU, while the outermost, and much more massive, planet is at about six times that distance; the fourth planet in the system has a nearly circular orbit; and the major axes of planets 1, 2, and 5 are currently nearly aligned, that is they have nearly equal $\varpi$ values.

\begin{figure*}
\includegraphics[width=\textwidth]{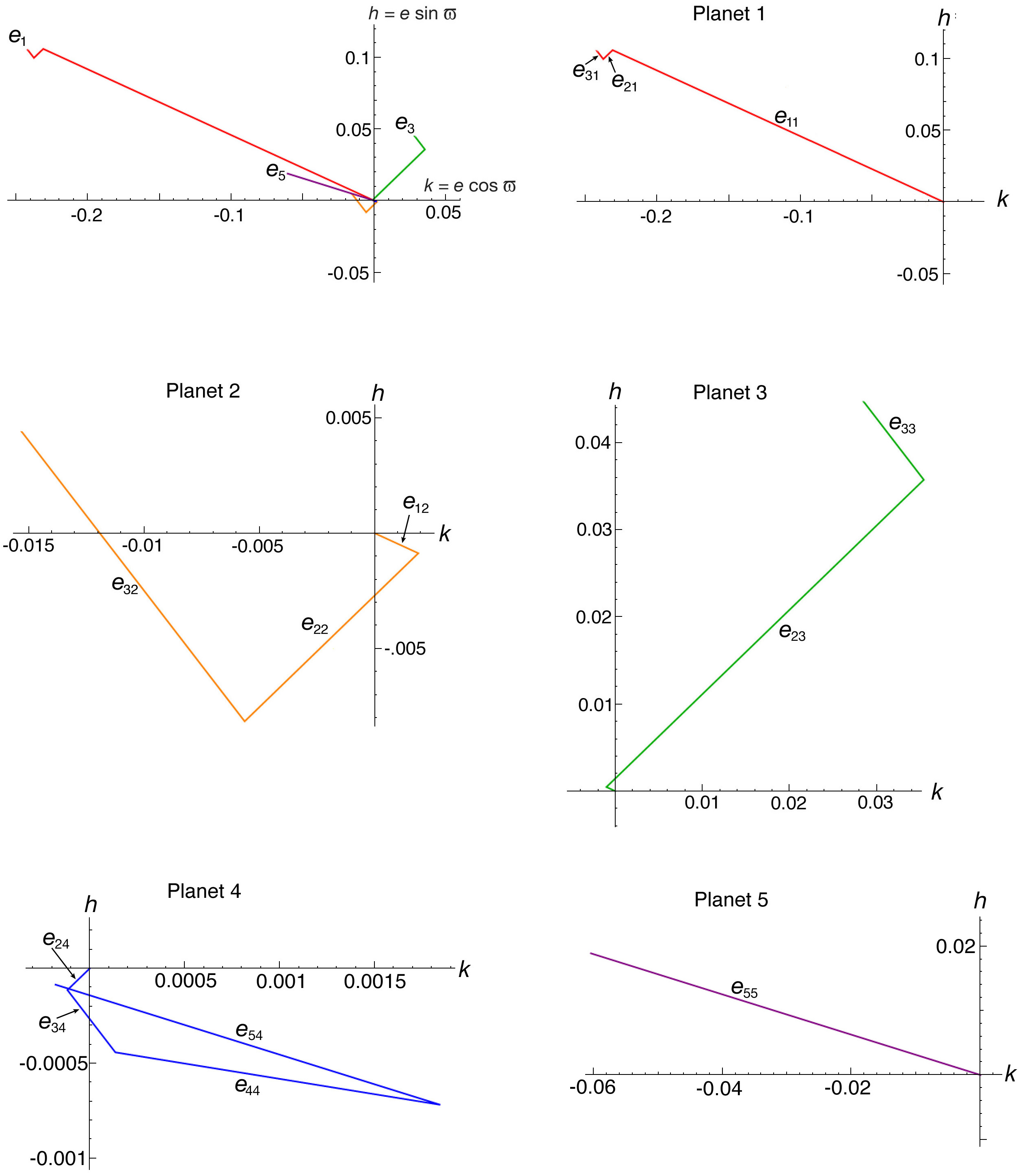}
\caption{
The eccentricity vectors ($k$,$h$) for each of the planets of 55 Cancri shown as a vector sum of the contributions from the five eigenmodes based on the Fischer et al. (2008)\cite{Fischer08} orbit solution (Table \ref{F1}). As plotted here, mode 1 starts from the origin, mode 2 starts from the end of mode 1, etc. The planets are colored from hotter to cooler colors in order of increasing semi-major axis: red, orange, green, blue, and purple for planets 1$-$5, respectively. Planets 1, 2, and 5 are all dominated by different eigenmodes. Remember that the contributions from a given eigenmode must be parallel or anti-parallel (i.e., a 180 degree difference). Top left: the ($k$,$h$) plots for all planets combined for comparison on a common scale. Other panels show the ($k$,$h$) plots for individual planets.
}
\label{figF}
\end{figure*}

The eigenmodes derived from the masses and semi-major axis values according to the theory in Sect.~\ref{sec2pt1} are shown in Table~\ref{F2}. The damping rate for each eigenmode is also shown (as a fraction of $F$) in Table \ref{F2}, according to the partitioning of the damping $F$ given by the theory of Sect. \ref{sec2pt2}. The actual magnitude ($E_m$) and phase ($\delta_m$) of each eigenmode is computed from the current $e$ and $\varpi$ values. From those results, Figure~\ref{figF} shows the current eccentricity vectors for planets 1 through 5, each as a vector sum of its 5 eigenmode components. Remember, for each eigenmode, the corresponding vector component circulates at the rate $g_m$ given in Table~\ref{F2}.

The alignment of the current e vectors for planets 1, 2, and 5 is evident in Figure~\ref{figF}. However, we note that planet 1's $e$ vector is dominated by the 1st eigenmode and planet 5's $e$ vector is dominated by the 5th eigenmode. For planet 2, the $e$ vector is predominantly composed of components from eigenmodes 2 and 3. Because the vector components for each eigenmode rotate at a different rate ($g_m$) from the other eigenmodes, the current alignment of these three orbits appears to be a coincidence (if not an observational artifact).

The alignment is not the result of damping to a quasi-fixed point. As discussed in Sect.~\ref{sec2pt2}, if the alignment had been due to eccentricity damping, all but the longest-lived eigenmode would have already damped out. Clearly this is not the case. In fact, the mode that should damp fastest (mode 1 as shown in Table~\ref{F2}) is actually the strongest in the sense of having the greatest amplitude $E_m$ in this system. These arguments show how secular theory can help determine whether or not a system may have experienced tidal evolution.

Another feature of this system is also revealed by consideration of how strongly each planet is affected by each eigenmode. From Table~\ref{F2} and Figure~\ref{figF}, one can see which planets are controlled by which modes. Specifically: planet 1 is most strongly affected by mode 1, with some contributions by modes 2 and 3; planet 2 is controlled by modes 2 and 3; planet 3 is dominated by mode 2, with an appreciable contribution from mode 3; planet 4 is controlled by modes 4 and 5, with a small contribution from mode 3; and planet 5 is strongly dominated by mode 5. Thus, the eigenmodes that affect the inner three planets do not strongly affect the outer two planets and \emph{vice versa}.

In other words, there are two distinct dynamical groups in the system. In physical terms, the inner three planets exchange angular momentum among themselves, but do not share very much with the outer two planets. Similarly the outer two planets exchange angular momentum predominantly between themselves. In general, the grouping of planets into subsets that share dominant eigenmodes is diagnostic of groups that exchange angular momentum among themselves due to mutual perturbations. An extreme case would be a system with negligible exchanges of angular momentum, in which case each planet would be controlled by only one eigenmode, and each eigenmode is associated with the behavior of only that one planet.

The dynamical grouping in this system is not what one might have anticipated based on the planets' semi-major axes alone. Because the orbit of planet 5 is much larger than that of any of the others, one might have thought that the inner four planets would be grouped together and the outermost planet would be on its own. But secular theory shows that the dynamical grouping is quite different. This result demonstrates the value of secular theory for helping to characterize the degree of dynamical packing of any system (as long as there are no mean-motion resonances involved).

Additional insight into the two dynamical groups in this system can be gained by further considering the structure of the eigenvectors. For the inner three planets, we see from Table~\ref{F2} that the second and third eigenmodes have significant values for all of the inner three planets, corresponding to the strong dynamical grouping. However, the first mode is quite concentrated in planet 1. Thus mode 1 provides an additional contribution to planet 1's eccentricity vector, but the link to the other members of this group is only through modes 2 and 3, reflecting the fact that planet 1 is too small to affect the other planets significantly. Accordingly, the damping rate for mode 1 is 0.996$F$ (from Table~\ref{F2}) nearly as fast as the direct damping $F$ (as defined in Eqn.~\ref{edampeqn}) of the eccentricity of planet 1, while the other modes damp far more slowly.

In the outer group, the coupling is primarily through the shared eigenmode 5, which is predominantly associated with the outer planet. Mode 5 reflects the effect of planet 5 on planet 4. Mode 4 is essentially only associated with planet 4; the two outer planets are not significantly coupled in this mode, reflecting the small mass of planet 4 and its lack of influence on Planet 5.

One other characteristic of the secular behavior (assuming the small value of $e_4$ in this orbit fit was correct) is that the initial conditions of the system (which set up the $E_m$ values) were apparently such that planet 4 periodically returns to a circular orbit. One way to explain this type of behavior is for planet-planet scattering to have occurred early in the system's history (e.g. \citealt{FordRasio05,Malhotra02,BarnesGreen07}), an event that set the the initial conditions for the subsequent secular interactions. If planet 4 were not affected by such interactions and remained on an initially circular orbit, while the other planets were stirred up by a scattering event, then the periodic secular behavior would have to be such that planet 4 periodically returns to a circular orbit. This type of on-going secular behavior could be suggestive of a long-ago epoch of planet-planet scattering. However, in this case, the scenario would have required simultaneous scattering of the inner planets (1, 2, and 3) and the outer one (5) but not planet 4, which seems improbable.


\subsection{Second Orbit Solution: Best Fit with Aliasing Corrections}
\label{55CncD10bf}

\begin{figure*}
\includegraphics[width=\textwidth]{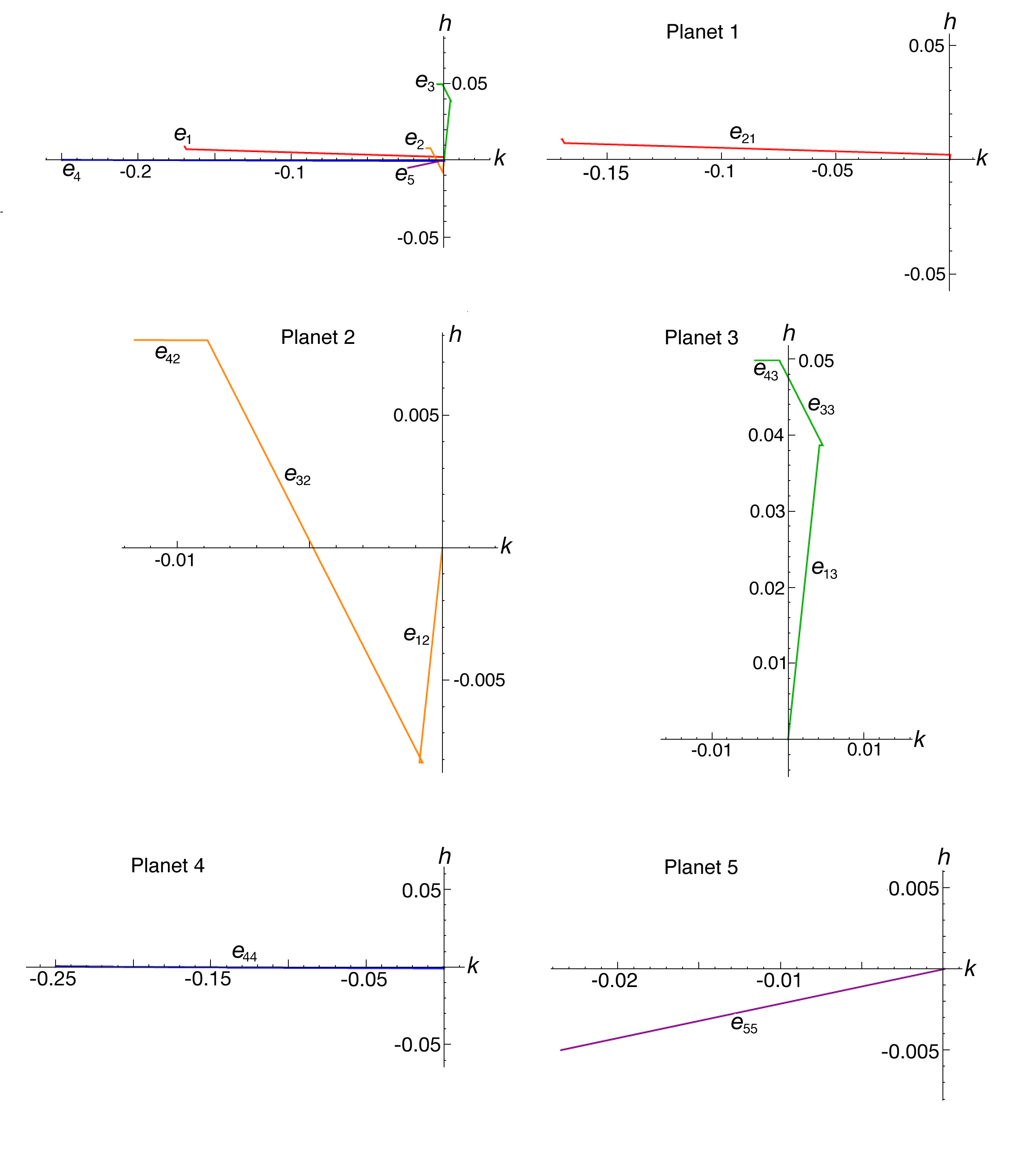}
\caption{
The eccentricity vectors ($k$,$h$) for the planets of 55 Cancri shown as a vector sum of the contributions from the five eigenmodes based on the best fit orbit solution by Dawson and Fabrycky 2010 (Tables~\ref{DF1} and \ref{DF2}). The planets are colored from hotter to cooler colors in order of increasing semi-major axis. Again, it is obvious that planets 1, 2, 4, and 5 are dominated by different eigenmodes.
}
\label{figDFbf}
\end{figure*}

\citet{DawsonFabrycky10} reconsidered the orbit solution based on data from \citet{Fischer08} and more recent observations, taking aliasing effects into consideration. Their best-fit orbital solution (our Table~\ref{DF1}), differs from that of Fischer et al. in several ways: the innermost planet is on an orbit much closer to its star; $e_4$ is much larger; and four of the planets' major axes are currently nearly aligned. Our secular analysis yields the eigenmodes for this solution as shown in Table~\ref{DF2}. Figure~\ref{figDFbf} shows each of the planets' $e$ vectors as a vector sum of the contributions from the five eigenmodes (c.f. Figure~\ref{figF}).

In this system, planets 1, 2, 4, and 5 are all controlled by different eigenmodes. So the current alignment of major axes is fortuitous, depending on the epoch of observations, as it was for the \citet{Fischer08} orbits. Also, the eigenmode whose eigenvector is most closely associated with the inner planet (here called mode 2) should damp fastest, yet it has a substantial $E_m$, indicating minimal eccentricity damping (just as for the solution by \citealt{Fischer08}). Furthermore, the small current amplitudes of eigenmodes 1 and 3 cannot be due to eccentricity damping, because the timescales for their damping are $\sim10^4$ times as long as for the undamped mode 2.  So the fact that modes 1 and 3 are not very excited is accidental (i.e. dependent on initial conditions). We emphasize that these conclusions apply only to the model system described by Dawson and Fabrycky's best-fit case.

Inspection of the eigenvectors and their amplitudes reveals similar dynamical groups in Dawson and Fabrycky's best fit to those found for Fischer et al.'s solution. However there are some key differences. First, planet 1 is not nearly as strongly coupled to the other inner planets. In the eigenvectors $V_{mp}$ derived from the Fischer et al. orbits the modes that dominated planets 2 and 3 (modes 2 and 3 in Table~\ref{F1}) also had significant effects on the innermost planet. In contrast, the equivalent modes derived from Dawson and Fabrycky's best fit (modes 1 and 3 in Table~\ref{DF1}) barely affect planet 1. This difference reflects the innermost planet's smaller semi-major axis and a consequent relative decoupling from the other planets. (Note, too, that given our convention for ordering the modes, the smaller semi-major axis caused a change in the numbering.)

Another difference is that the 4th eigenmode plays a much greater role in the inner planets' eccentricities in the solution by Dawson and Fabrycky, despite the fact that the normalized eigenvector $V_{4p}$ is nearly the same in both solutions. It has a much greater magnitude ($E_4$ is $\sim$150 times larger) following from the larger value of $e_4$ and thus both planets 2 and 3 receive noticeable contributions from the 4th eigenmode.

This result demonstrates that two factors influence what constitutes a dynamical grouping of planets: (1) the distribution of $V_{mp}$ values for one or more modes affecting those planets and (2) the magnitude of the eigenmodes, as both $V_{mp}$ and $E_m$ affect the degree to which various planets are exchanging angular momentum.


\subsection{Third Orbit Solution: $e_1$ Set to Zero}
\label{55CncD10alt}

\citet{DawsonFabrycky10} also derived an orbital fit for 55 Cancri in which planet 1's eccentricity was held to zero (Table~\ref{DF3}). This constraint was motivated by the idea that the eccentricity should have been damped by tides, given the proximity to the star and reasonable assumptions about tidal parameters. In fact, the exponential damping timescale (1/$F$) for the inner planet is very short.  Assuming tidal parameters $Q_p$ = 500 and $Q_*$ = $3\times10^5$, and a radius of 2$R_{Earth}$ \citep{Winn11}, with conventional formulae for tidal damping (e.g. Jackson et al. (2008)\cite{Jackson08}) yields 1/$F$ = 50,000 years, so substantial damping is to be expected.  (In this case, the accompanying change in semi-major axis would not affect the eccentricity-damping timescale.)  Thus, it is certainly appropriate to seek an orbital fit that, unlike the ``best-fit'' solution, is consistent with tidal damping.

Dawson and Fabrycky's solution with the inner planet constrained to a circular orbit fit the data nearly as well as
their best fit, according to the chi-squared criterion. However, Dawson and Fabrycky expressed skepticism about this solution, realizing that that the influence of other planets in the system could pump up planet 1's eccentricity, in which case a low eccentricity would be short lived and therefore improbable.

\begin{figure*}
\includegraphics[width=\textwidth]{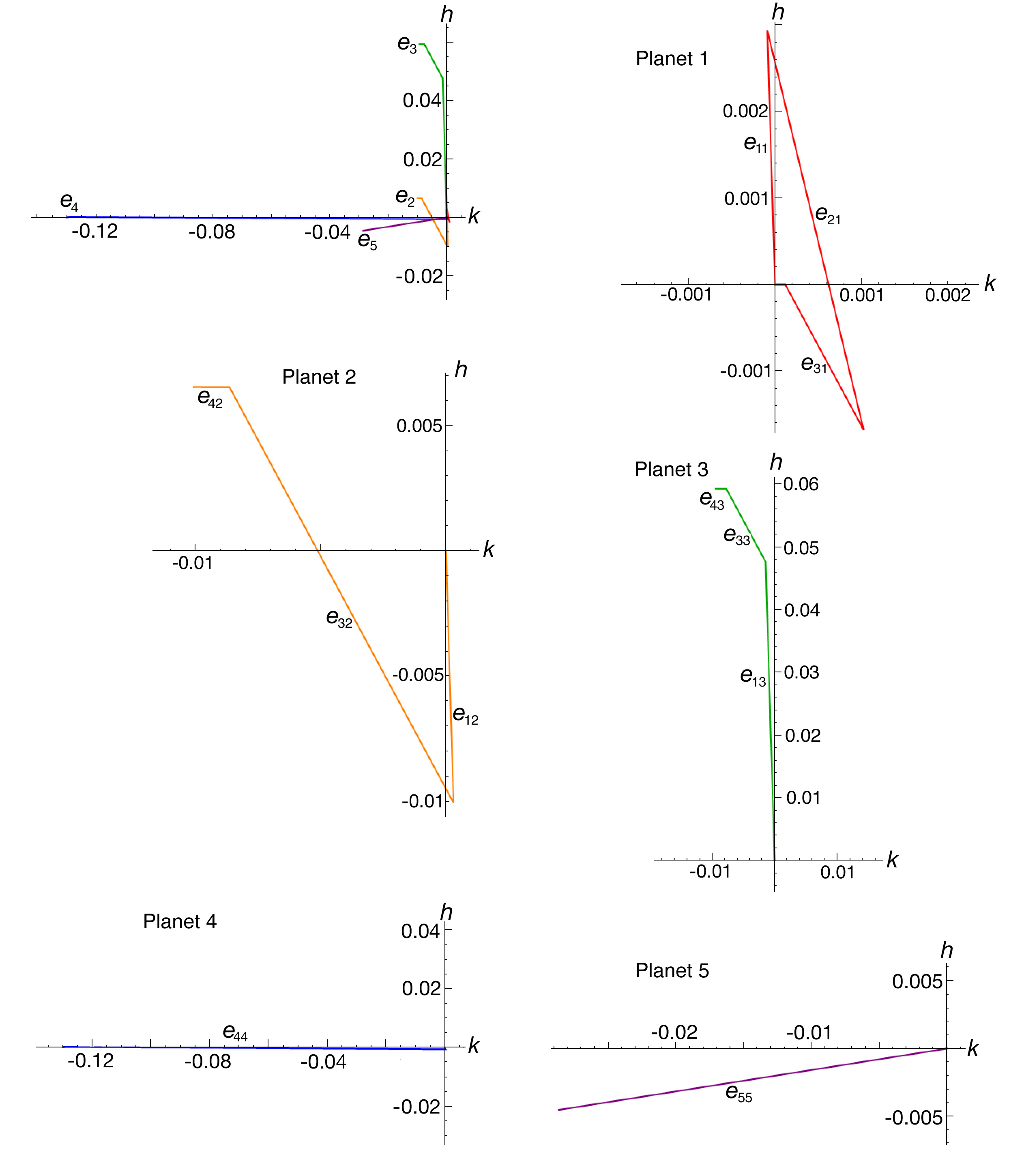}
\caption{
The eccentricity vectors ($k$,$h$) for the planets of 55 Cancri shown as a vector sum of the contributions from the five eigenmodes based on the alternative fit with $e_1=0$ by Dawson and Fabrycky 2010 (Tables~\ref{DF3} and \ref{DF4}). The planets are colored from hotter to cooler colors in order of increasing semi-major axis. Planets 2, 4, and 5 are still fortuitously currently near alignment. The overall qualitative character of the system has not changed from Table~\ref{DF2} and Figure~\ref{figDFbf} with the exception that planet 1 now will not cycle to a very high eccentricity.
}
\label{figDFe1zero}
\end{figure*}

Analysis of this system with secular theory (Table \ref{DF4} and Figure \ref{figDFe1zero}) quantifies that problem. It also demonstrates other reasons to reject this solution, and suggests a better approach to finding orbital fits consistent with the likelihood of past tidal evolution. Figure \ref{figDFe1zero} illustrates the concern raised by Dawson and Fabrycky about their solution:  Even though the current $e_1$ is constrained to zero, it is the sum of vectors that rotate in ($h$,$k$) space, and thus would increase periodically to values as large as ~0.01.  Table \ref{DF4} shows that the eigenvectors and eigenfrequencies are very similar to those from Dawson and Fabrycky's best fit (Table \ref{DF2}).

The only significant difference is found in the eigenmode amplitudes.  Most significantly, the magnitude $E_2$ of mode 2 is much less than in the best-fit solution, because this mode is almost exclusively associated with the inner planet, whose eccentricity is now constrained to zero. Indeed, according to Table \ref{DF4}, the damping timescale for mode 2 should be almost identical to 1/$F$, so the fact that its amplitude (0.0048) is much smaller than it was in the best-fit solution (0.17) is an improvement.  However this amplitude is still too large to be consistent with tidal damping, because it should have decreased below 0.0048 in $<10^5$yr, that is only 5 times the short timescale 1/$F$.  Similarly, the amplitudes of modes 1 and 3 are too large to be consistent with the tidal damping of the inner planet.  These two modes should damp on timescales $10^4$ times longer than 1/$F$, which is still an order of magnitude shorter than the age of the system.  Hence the current substantial amplitudes of these modes are inconsistent with tidal damping.

In systems like those derived in Dawson and Fabrycky's solutions, tidal damping of the inner planet would have quickly eliminated mode 2, then 10,000 times more slowly eliminated modes 1 and 3, while modes 4 and 5 would remain long-lived.

The secular analysis thus suggests a more meaningful way to account for tidal evolution in solutions for orbital parameters from observations of extrasolar multi-planet systems.  Rather than constrain the damped planet's orbit to circular, one should look for solutions where the current amplitudes of the faster damping modes (those that are expected to have damped away over the age of the system) are near zero.  This constraint will, of necessity, require additional iterative searches for viable current orbits, but the resulting solutions will not be inconsistent with tidal evolution.

For densely packed systems, another consideration that may be used to constrain a planetary system's architecture as well as its tidal evolution is the effect of tidal damping on orbital stability.  As the amplitudes of the secular modes vary, they may affect the stability of this system.  Application of these considerations has been pioneered by Lovis et al. (2011)\cite{Lovis11} in the context of the HD 10180 system, for which there is some evidence of as many as seven planets.


\section{Discussion}
\label{discussion}

As shown in this paper, classical secular theory is a useful tool that yields descriptive parameters for characterizing planetary systems in ways that can elucidate current dynamical interactions as well as illuminate and constrain the history of a given system. The method has its limitations of course.  As noted above, numerical integrations can yield far more precise ephemerides, given current orbits.  On the other hand, where observations are limited, the precision of secular theory could be adequate if commensurate with uncertainties in the observed orbits.  The classical theory considered here is also limited by the assumption that there are no significant mean-motion resonances involved.  This constraint will limit its applicability in considerations of long-term evolution, during which any system may well pass through resonant conditions.

Another limitation is that classical theory as discussed here ignores all terms in the mutual disturbing functions higher than second order in the eccentricities.  This limitation is significant because many extra-solar planets have large eccentricities, some approaching unity. For eccentricities greater than $\sim$0.4, qualitatively different modes of behavior can come into play (e.g. \citealt{MichMal04}).

On the other hand, classical secular theory can provide surprisingly good qualitative descriptions of the character of the actual behavior for eccentricities up to $\sim$0.4. For example, the behavior of the hypothetical system studied by Mardling (2007), as well as its evolution when one orbit’s eccentricity is damped, fits what would be expected from classical theory (see Section \ref{eccdamp} below).  Thus, if the objective is a physical understanding of the processes that govern the planetary interactions and evolution, the classical method has some advantages, as long as its limitations are also borne in mind.  Some of the ways that this approach can be valuable are summarized below.


\subsection{Dynamical Groups}
\label{dyngroup}

In each of the examples discussed here, inspection of the relative role of each eigenmode in the behavior of each planet helps characterize how closely the planets are dynamically linked.  At one extreme, if each eigenvector had a significant component for only one planet, the planets would be in effect dynamically isolated, the opposite extreme from a ``packed'' planetary system \citep{BarnesQuinn04,BarnesRaymond04,RaymondBarnes05,Raymond06,BarnesGreen06b}).  The dynamical groups that can be inferred from the sharing of eigenmodes 
represent the pathways through which planets can exchange angular momentum. They are an indication of the degree of planet-planet interaction, and hence of the denseness of planetary packing.

Strongly coupled groups imply highly packed systems, and \emph{vice versa}.  Where a system divides into separate groups of planets only linked among themselves, there is a dynamical gap with possible implications about the completeness of either planet formation or detection.  The fact that most planets considered here are secularly coupled supports the notion that planets tend to form in fairly compact configurations.

Nevertheless, we have seen that subsets of planets often be dynamically isolated from others.  One example was the system described by Fischer et al.'s (2008)\cite{Fischer08} solution for 55 Cancri which divided into two groups: the inner three planets and the outer two planets.  In contrast, the solutions for the same system by \citet{DawsonFabrycky10} has the inner planet relatively uncoupled from the rest.

Interestingly, the fact that a set of planets are dynamically linked through a common eigenmode does not necessarily mean that they exchange angular momentum. To exchange angular momentum, the eccentricities of those planets must involve more than one eigenmode so that the contributions from the modes can periodically add constructively and destructively. For a set of planets whose eccentricities are governed by only one eigenmode (i.e. where the other modes have zero amplitudes), all the eccentricity values would be fixed and there would be no exchange of angular momentum.


\subsection{Effects of Eccentricity Damping}
\label{eccdamp}

If any process acts directly to damp the eccentricity of one planet, secular coupling with the other planets will distribute its effects. Thus the magnitudes of all of the eigenmodes of the system will decrease, indirectly damping the eccentricities of all the planets.  How fast each eigenmode damps reflects how concentrated (or not) the eigenvector is in the directly affected planet.  The damping process that has recieved the most attention is tidal dissipation, which acts most strongly on the inner planet in most systems.  Tides also affect semi-major axes, which in turn can have an effect on the rates of damping of the various components of each eigenvector, as first shown by \citet{WuGoldreich02} for a special case with only one non-zero eigenmode amplitude, and more generally by \citet{GrVLtides}. 

Here though we have only considered the effect of eccentricity damping on the inner planet, and shown how the modification of secular interactions can result in indirect damping of all the orbits. The sequential damping of the eigenmodes of the classical secular theory can be surprisingly accurate even when eccentricities might seem to be outside the expected range of validity of the small-eccentricity approximation.

An example of classical secular theory working at moderate eccentricities and of how the influence of tides manifests is found in comparing with \citet{Mardling07} in which she develops a secular model that allows for large ecentricities as well as tidal evolution.
This theory does have some significant restrictions:  It applies only to a system of two planets; The ratio of semimajor axes (inner/outer) must be small; The solution is in a form that requires numerical integration with an artificially low $Q$ value.  Nevertheless, this approach extends secular theory to conditions of large eccentricity where the classical theory would be quantitatively unreliable.

Mardling applied the theory to a hypothetical system of two planets with fairly large $e$ values (0.1 and 0.4) and described an evolutionary sequence that went through three stages, illustrated in Mardling's Figures 3 and 4: First, the pericenter longitudes circulate relative to one another; then the system transitions to a stage at which the pericenter longitudes librate relative to one another with a gradual decrease in the amplitude; Finally, after the libration has damped down, the pericenters are locked in alignment and the eccentricities' values are locked in a fixed ratio as their values gradually decrease (i.e. the quasi-fixed-point state).

\begin{figure*}
\includegraphics[width=\textwidth]{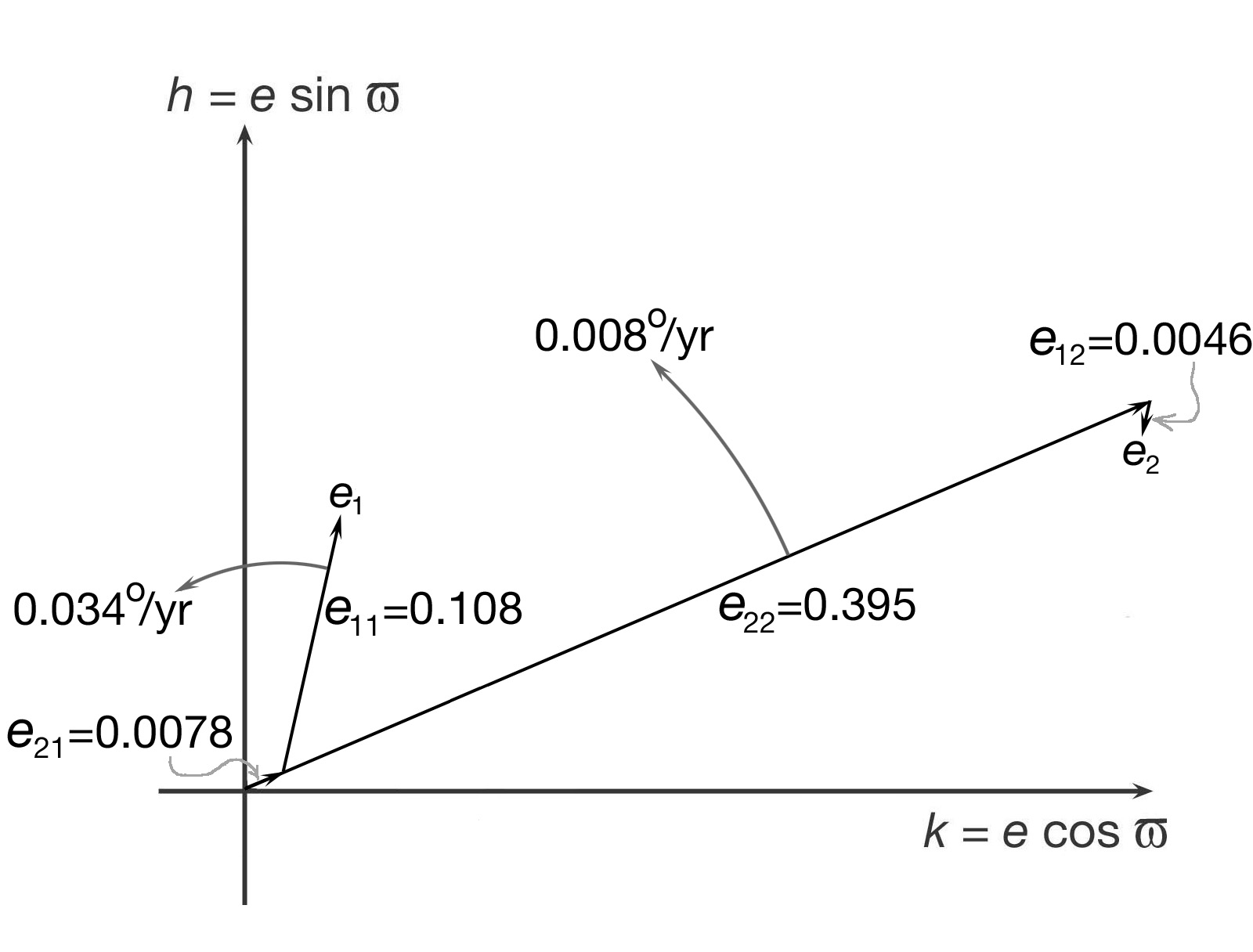}
\caption{
With the parameters and initial conditions of the hypothetical system investigated by Mardling (2007)\cite{Mardling07}, classical secular theory yields the behavior of the system illustrated here.  The eccentricity $e_1$ of the inner planet is given by the vector sum in ($h$,$k$) space of components $e_{11}$ from eigenmode 1 and $e_{21}$ from eigenmode 2, with values shown.  Similarly, the eccentricity $e_2$ of the outer planet is the sum of components $e_{12}$ (always directed opposite $e_{11}$) and $e_{22}$ (aligned with $e_{11}$).  As $e_{11}$ circulates at the rate 0.026$^o$/yr relative to the mode 2 components, $\varpi_1 - \varpi_2$ circulates through 360$^o$. Later as mode 1 damps down (so $e_{11} < e_{21}$), $\varpi_1 - \varpi_2$ librates.  The behavior and evolution closely follows that derived by Mardling.
}
\label{figM}
\end{figure*}

Given that one planet starts with $e$=0.4 in this system, the classical theory cannot be expected to give a quantitatively precise description of this behavior.  However, we can explore the limits of its validity, by comparing its description of this system (Figure~\ref{figM}) with the more precise solution by Mardling.  The behavior shown in Figure~\ref{figM} is very similar to that found by Mardling in the initial phase of the evolution.  Our classical theory shows that mode 1 damps $10^3$ times faster than mode 2. The decrease of the mode 1 components ($e_{11}$ and $e_{12}$ in Figure~\ref{figM}) then reproduces remarkably accurately the behavior described by Mardling, with the transition to libration after $e_{11}$ becomes smaller than $e_{21}$, and then transition to the quasi-fixed point after mode 1 has effectively died away.

Thus, while Mardling's method provides a more precise result for this system, the classical approach provides a surprisingly good description of the behavior, and complements it in several ways. Besides applying to cases where Mardling's restrictions may not be valid, it offers a more straightforward physical interpretation:  The three-stage process described by Mardling reduces to simply the gradual damping of one of the eigenmodes. 

A system that has undergone considerable direct damping of one (or more) planet's eccentricity will eventually have only one remaining non-zero eigenmode amplitude, in which case there is an alignment of major axes (the quasi-fixed-point condition). However, an alignment of major axes does not necessarily imply that damping has taken place. We have seen that solutions for 55 Cancri (e.g. Section \ref{55CncF08} and \ref{55CncD10bf}) show some currently aligned orbits, but these are simply fortuitous alignments of the dominant eigenmodes for those planets at the present time.  As the eccentricity vectors circulate over secular timescales, these temporary configurations would vanish. Another example is the $\mu$ Arae system described by \citet{Pepe07}, in which major axes are currently aligned. Even in systems where the eigenvectors are such that alignments among some of the orbits endure over secular timescales, this may not be evidence for past damping. For example, in the 61 Virginis system an alignment had been interpreted as an indication of tidal evolution \citet{Vogt61Vir}, but secular analysis by \citet{GrVL61Vir} shows that all of the eigenmodes still retain substantial magnitudes so any damping has been minimal. As discussed by \citet{GrVL61Vir}, for 61 Virginis there may even be a propensity toward alignment due to the particular semi-major axis values and masses in the system. Alternative explanations for the fact that so many extra-solar planetary orbits are aligned may be that there is some as yet unidentified tendency of the orbit-fitting process to introduce such apparent alignments as a systematic error, or that, for systems with lasting (not fortuitious) alignment, this may help stabilize a system by minimizing close approaches.

While the quasi-fixed-point solution is the expected final stage of the damping process, this may not always be the case.  There is no reason why the damping rates of the longest lived eigenmodes are necessarily very different.  In some systems, two or more eigenmodes may damp out on similar timescales, such that there is no single longest-lived mode.  In that case, there would never be a quasi-fixed-state (i.e. aligned axes) condition since these eigenmodes will die out together rather than in sequence.  Also, if one or more of the outer planets are only weakly coupled to the inner planet then there may be several modes which would not damp appreciably even on timescales as long as the age of the universe or longer. For example, in the 55 Cancri system the 4th and 5th eigenmodes damp much much slower than the other modes and will both therefore persist long term.


\subsection{Variation of Eccentricities due to Secular Interaction}
\label{eccvari}

Secular theory can help determine whether an eccentricity that is determined (or assumed) to be small will stay that way over secular timescales. In \citet{DawsonFabrycky10} orbit-fit solution for 55 Cancri with $e_1$ held at zero (Sect.~\ref{55CncD10alt}), secular theory allows us to quantify the periodic increase in $e_1$'s value.  We found that the increase was small, so the assumption of small $e_1$ was reasonably self-consistent.  However, because the circular-orbit constraint was motivated by the expectation that the inner planet would have undergone considerable direct damping, a better approach to such a case in the future will be to find a fit in which the shorter-lived eigenmodes have damped away.  Such a solution would require new techniques, perhaps an iterative process, but would provide a more well-founded constraint on an orbital fit to account for the likelihood of past eccentricity damping in a system with such a close-in planet.

Another interesting example is the system around Gliese 581 \citep{Bonfils05,Udry07,VogtGJ581}. \citet{VogtGJ581} reported that the system appeared to contain one planet (g) within the star's habitable zone (as conventionally defined by the potential for liquid water on a planet's surface). While Vogt et al. deemed gleaning eccentricities from the data too uncertain, \citet{AngEsc11} showed that if planet g exists, then its outer neighbor, d, likely has an eccentricity $\sim$0.1. Our analysis of the eigenvectors, based on classical secular theory, shows that if $e_d\sim0.1$ then $e_g$ must also periodically get approximately as large. That is, if planet g exists, the variation in insolation over each orbit brings into question its habitability, even though its semi-major axis places it inside the habitable zone. The system of Vogt et al. has more recently been called into question by \citet{Tuomi11} and \citet{Greggory11}, but the example nevertheless illustrates another way that classical secular theory can be useful.

The examples demonstrate the value of applying classical secular theory to assess the variability of eccentricities under perturbations by other planets. Even under circumstances where classical secular theory is not precise (e.g. at high eccentricities), it can provide a basis for identifying systems worthy of further examination by more precise methods.


\subsection{Planet-planet Scattering}
\label{nearsep}

For two planet systems, classical secular theory has also provided a way to identify the signature of planet-planet scattering that likely has occurred early in the history of planetary systems, and which may play a major role in setting up their current architecture \citep{RasioFord96}. \citep{BarnesGreen06a} (see also \citealt{BarnesQuinn04}) showed that the type of scenario demonstrated by \citet{FordRasio05} produces systems whose secular behavior lies near a boundary between libration and circulation, or more generally where one or more planets periodically return to near-circular orbits. Such systems are very common \citep{BarnesGreen06c,BarnesGreen07}, so classical secular theory provides a useful tool for identifying candidate systems that may have been influenced by this crucial process.

This type of behavior is not necessarily the signature of planet-planet scattering in all cases however. Recall for example the 4th planet in the 55 Cnc system as described by \citet{Fischer08} (Sect.~\ref{55CncF08}). As discussed in Sect.~\ref{55CncF08}, it appears difficult to arrange a planet-planet scattering event that leaves only one planet in the middle of the planetary system to exhibit periodic returns to a circular orbit. Because of such complications, such secular behavior in systems of three or more planets could be used as a way to screen for systems that might have undergone planet-planet scattering, but it could not be considered a definitive signature of such an event.


\section{Conclusion}
\label{concl}

We conclude that classical secular theory is a powerful tool for interpreting the observed dynamical state of a planetary system.  Numerical integration is a much more precise way to simulate the on-going behavior of a system, but each integration represents in a sense an anecdotal case.  Analytical theory provides a qualitative characterization and classification that allows interpretation and yields implications for the formation and history of these systems. Classical secular theory thus provides a valuable complement to other approaches to understanding the dynamics of planetary systems.


\begin{acknowledgements}
We thank D. Fabrycky, D. Hamilton, R. Barnes, R. Dawson, and B. Jackson for their insightful discussions on this topic. A. Morbidelli provided essential help with the manuscript. This work was partially funded by the National Science and Engineering Research Council of Canada through a Postgraduate Scholarship - Masters.
\end{acknowledgements}



\clearpage

\begin{table*}
\caption{The orbits of the planets in 55 Cancri as determined by Fischer et al. 2008\cite{Fischer08}, their Table 4.}
\label{F1}
\begin{tabular}{c|cccc}
\hline\noalign{\smallskip}
Planet & $M\sin(i)$ ($M_J$) & $a$ (AU) & $e$ & $\varpi (^o)$  \\
\noalign{\smallskip}\hline\noalign{\smallskip}
1 (e) & 0.0241 	& 0.038 	& 0.2637 	& 156.5 \\
2 (b) & 0.8358 	& 0.115 	& 0.0159 	& 164.0 \\
3 (c) & 0.1691 	& 0.241 	& 0.0530 	& 57.4 \\
4 (f) & 0.1444 	& 0.785 	& 0.0002 	& 205.6 \\
5 (d) & 3.9231 	& 5.901 	& 0.0633 	& 156.6 \\
\noalign{\smallskip}\hline
\end{tabular}
\end{table*}

\begin{table*}
\caption{Values pertaining to the eigenmodes of the 55 Cancri system using the orbits found in Table \ref{F1} above, Fischer et al.'s (2008)\cite{Fischer08} best fit. The eigenvectors ($V_{mp}$), eigenfrequencies ($g_m$), and damping rates are all determined using only the masses and semi-major axes of the planets. The "damping rate" is that due to eccentricity damping and is given in terms of $F$, the direct damping as defined in Eqn.~\ref{edampeqn}. The amplitudes of the eigenmodes ($E_m$) are determined using the current eccentricities and longitudes of pericenter ($\varpi$).}
\label{F2}
\begin{tabular}{@{\hspace{5pt}}c@{\hspace{5pt}}|c@{\hspace{5pt}}c@{\hspace{5pt}}c@{\hspace{5pt}}c@{\hspace{5pt}}c|c@{\hspace{5pt}}c@{\hspace{5pt}}c}
\noalign{\smallskip}\hline\noalign{\smallskip}
\multirow{3}{*}{Mode} & \multicolumn{5}{|c|}{$V_{mp}$} & \multirow{3}{*}{$g_m (^o/yr)$} &
	\multirow{3}{*}{$\frac{1}{E_m}\frac{dE_m}{dt}$} & \multirow{3}{*}{$E_m$}  \\
\cline{2-6}
 & \multicolumn{5}{|c|}{Planet} & & & \\
 & 1 (e) & 2 (b) & 3 (c) & 4 (f) & 5 (d) & & & \\
\noalign{\smallskip}\hline\noalign{\smallskip}
1 & 0.99996 	& -0.0082 	& 0.0044 	& -2$\times10^{-6}$ 	&  -9$\times10^{-11}$ 	& 1.4 		& -0.996$F$ 				& 0.25 \\
2 & -0.17 	& -0.20 		& 0.96 		& -0.0031 			& -2$\times10^{-7}$ 		& 0.81 		& -0.0016$F$ 			& 0.053 \\
3 & 0.34 	& 0.77 		& 0.55 		& -0.020 			& -6$\times10^{-7}$ 		& 0.15 		& -0.0028$F$ 			& 0.021 \\
4 & -0.0038 	& -0.0093 	& -0.012 	& -0.9999 			& 0.00045				& 0.019 		& -5$\times10^{-7}F$ 	& 0.0017\\
5 & 0.00013 	& 0.00033 	& 0.00043 	& 0.034 				& 0.9994 				& 0.000062 	& -8$\times10^{-12}F$ 	& 0.063 \\
\noalign{\smallskip}\hline\noalign{\smallskip}
\end{tabular}
\end{table*}

\begin{table*}
\caption{The best fit orbits of the planets of 55 Cancri as found by Dawson and Fabrycky (2010)\cite{DawsonFabrycky10}, their Table 7.}
\label{DF1}
\begin{tabular}{c|cccc}
\hline\noalign{\smallskip}
Planet & $M\sin(i)$ ($M_J$) & $a$ (AU) & $e$ & $\varpi (^o)$  \\
\noalign{\smallskip}\hline\noalign{\smallskip}
1 (e) & 0.0261 	& 0.01564 	& 0.17 	& 177 \\
2 (b) & 0.826 	& 0.1148 	& 0.014 	& 146 \\
3 (c) & 0.171 	& 0.2403 	& 0.05 	& 95 \\
4 (f) & 0.150 	& 0.781 		& 0.25 	& 180 \\
5 (d) & 3.83 		& 5.77 		& 0.024 	& 192 \\
\noalign{\smallskip}\hline
\end{tabular}
\end{table*}

\begin{table*}
\caption{Values pertaining to the eigenmodes of the 55 Cancri system using the orbits found in Table \ref{DF1} above, Dawson and Fabrycky's (2010)\cite{DawsonFabrycky10} best fit scenario (c.f. Table~\ref{F2}).}
\label{DF2}
\begin{tabular}{@{\hspace{5pt}}c@{\hspace{5pt}}|c@{\hspace{5pt}}c@{\hspace{5pt}}c@{\hspace{5pt}}c@{\hspace{5pt}}c|c@{\hspace{5pt}}c@{\hspace{5pt}}c}
\noalign{\smallskip}\hline\noalign{\smallskip}
\multirow{3}{*}{Mode} & \multicolumn{5}{|c|}{$V_{mp}$} & \multirow{3}{*}{$g_m (^o/yr)$} &
	\multirow{3}{*}{$\frac{\dot{E}_m}{E_m}$} & \multirow{3}{*}{$E_m$}  \\
\cline{2-6}
 & \multicolumn{5}{|c|}{Planet} & & & \\
 & 1 (e) & 2 (b) & 3 (c) & 4 (f) & 5 (d) & & & \\ 
\noalign{\smallskip}\hline\noalign{\smallskip}
1 & -0.053 		& 0.20 		& -0.98 		& 0.0033 	& 3$\times10^{-7}$ 	&  0.81 		& -0.000098$F$ 		& 0.040 \\
2 & -0.999996 	& 0.00069 	& 0.0026 	& -0.000014 & -1$\times10^{-9}$ 	& 0.62 		& -0.9998$F$ 		& 0.17 \\
3 & 0.087 		& 0.82 		& 0.57 		& -0.024 	& -6$\times10^{-7}$ 	& 0.14 		& -0.0001$1F$ 		& 0.022 \\
4 & -0.00095 	& -0.011 	& -0.013 	& -0.9998 	& 0.00050 			& 0.019 		& -2$\times10^{-8}$ 	& 0.25 \\
5 & 0.000033 	& 0.00040 	& 0.00049 	& 0.035 		& 0.9994 			& 0.000068 	& -4$\times10^{-13}$ & 0.024 \\
\noalign{\smallskip}\hline\noalign{\smallskip}
\end{tabular}
\end{table*}

\begin{table*}
\caption{An alternative fit with $e_1$=0 for the planets of 55 Cancri as found by Dawson and Fabrycky (2010)\cite{DawsonFabrycky10}, their Table 8.}
\label{DF3}
\begin{tabular}{c|cccc}
\hline\noalign{\smallskip}
Planet & $M\sin(i)$ ($M_J$) & $a$ (AU) & $e$ & $\varpi (^o)$  \\
\noalign{\smallskip}\hline\noalign{\smallskip}
1 (e) & 0.0258 	& 0.01564 	& 0.0 	& 0 \\
2 (b) & 0.825 	& 0.1148 	& 0.012 	& 147 \\
3 (c) & 0.172 	& 0.2402 	& 0.06 	& 99 \\
4 (f) & 0.150 	& 0.781 		& 0.13 	& 180 \\
5 (d) & 3.83 		& 5.77 		& 0.029 	& 189 \\
\noalign{\smallskip}\hline
\end{tabular}
\end{table*}

\begin{table*}
\caption{ Values pertaining to the eigenmodes of the 55 Cancri system using the orbits found in Table \ref{DF3} above, Dawson and Fabrycky's (2010)\cite{DawsonFabrycky10} alternative fit with $e_1=0$ (c.f. Table~\ref{F2}).}
\label{DF4}
\begin{tabular}{@{\hspace{5pt}}c@{\hspace{5pt}}|c@{\hspace{5pt}}c@{\hspace{5pt}}c@{\hspace{5pt}}c@{\hspace{5pt}}c|c@{\hspace{5pt}}c@{\hspace{5pt}}c}
\noalign{\smallskip}\hline\noalign{\smallskip}
\multirow{3}{*}{Mode} & \multicolumn{5}{|c|}{$V_{mp}$} & \multirow{3}{*}{$g_m (^o/yr)$} &
	\multirow{3}{*}{$\frac{1}{E_m}\frac{dE_m}{dt}$} & \multirow{3}{*}{$E_m$}  \\
\cline{2-6}
 & \multicolumn{5}{|c|}{Planet} & & & \\
 & 1 (e) & 2 (b) & 3 (c) & 4 (f) & 5 (d) & & & \\
\noalign{\smallskip}\hline\noalign{\smallskip}
1 & 0.060 		& -0.21 	& 0.98	& -0.0033 	& -3$\times10^{-7}$ 	& 0.77 		& -0.00013$F$ 		& 0.049 \\
2 & 0.999996 	& -0.00059 	& -0.0028	& 0.000014 	& 1$\times10^{-9}$ 	& 0.62 		& -0.9998$F$ 		& 0.0048 \\
3 & -0.083	 	& -0.82 	& -0.57 	& 0.024 	& 6$\times10^{-7}$ 	& 0.13		& -0.00010$F$ 		& 0.023 \\
4 & 0.00091	 	& 0.011 	& 0.013	 & 0.9998 	& -0.00050 			& 0.018 		& -2$\times10^{-8} F$ 	& 0.13\\
5 & 0.000032 	& 0.00040 	& 0.00049	& 0.035 	& 0.9994			& 0.000065 		& -4$\times10^{-13} F$ 	& 0.029\\
\noalign{\smallskip}\hline\noalign{\smallskip}
\end{tabular}
\end{table*}


\end{document}